\RequirePackage{fix-cm}
\documentclass[twocolumn,epjc3]{svjour3}  
\smartqed  
\RequirePackage{graphicx}
\usepackage{color}
\usepackage{amssymb,bbold}
\usepackage{graphicx,color}
\usepackage[usenames,dvipsnames]{xcolor}
\usepackage[section]{placeins}
\usepackage{amsmath}
\usepackage{subfigure}
\usepackage[normalem]{ulem}
\usepackage{booktabs}
\usepackage{xcolor}
\usepackage{epsfig,graphics,color,graphicx,amsmath}
\colorlet{darkgreen}{green!50!black}
\colorlet{brightyellow}{yellow!75!red}
\colorlet{orange}{red!50!yellow}
\colorlet{darkblue}{blue!60!black}
\colorlet{darkred}{red!80!black}
\usepackage{widetext}
\def\bwt{\begin{widetext}}
\def\ewt{\end{widetext}}

\usepackage{soul}

\usepackage{mathtools}
\usepackage{mathrsfs}
\usepackage{bm}
\usepackage{relsize}
\usepackage{indentfirst}

\usepackage{xcolor}

\def\psla{\slash \! \! \!}

\usepackage{amssymb,amsmath,multirow,epsfig,graphicx,color}

\usepackage{epsfig}
\usepackage{graphicx,amsmath}
\def\be{\begin{eqnarray} &&}

\def\ee{\end{eqnarray}}
\def\psla{\slash \! \!\! }

\begin{document}

\title{On the momentum space structure of the quark propagator}
\author{O Oliveira \thanksref{e1,addr1} \and T Frederico \thanksref{e2,addr2} \and W. de Paula \thanksref{e3,addr2}}
\institute{CFisUC, Departament of Physics, University of Coimbra, 3004-516 Coimbra, Portugal \label{addr1} \and
Instituto Tecnol\'ogico de Aeron\'autica,  DCTA,
12228-900 S\~ao Jos\'e dos Campos,~Brazil \label{addr2}}
\thankstext{e1}{e-mail: orlando@uc.pt} 
\thankstext{e2}{e-mail: tobias@ita.br}
\thankstext{e3}{e-mail: wayne@ita.br} 

\date{\today}
\maketitle
\begin{abstract}
The structure of the quark propagator in momentum space is explored taking into account non-perturbative 
QCD dynamics constraints for the quark spectral densities derived previously.
We assume that the scalar and vector component of the quark propagator share a  simple pole but not its residuum, together with other structures.
Furthermore, a connection between the poles of the quark propagator and the zeros of the quark wave function $Z(p^2)$ is 
established. 
Asymptotic scaling laws for the representation of the quark propagator, after removing the shared pole, are also derived. 
The confrontation of our results with lattice data for the full QCD quark propagator data are in good agreement. 
Exploring the link with the lattice data and looking at the Bethe-Salpeter vertex and amplitude, in the chiral limit, 
we are able to provide estimations for these quantities, for $f_\pi$ and for the shared pole mass. The pole mass reproduces
the constituent quark mass  used in the quark models.

\end{abstract}


\section{Introduction and Motivation}

In a quantum field theory the analytic structure of the propagators determines the properties of the corresponding
quanta. In QCD, having the propagators
can help understanding, among other hadronic properties, the confinement mechanism and the origin of dynamical chiral symmetry breaking~\cite{QCD50Y}. However, accessing the analytic structure of 
propagators without relying on perturbation theory  or without making further assumptions is rather difficult. For example, lattice field theory can deliver important 
information about the non-perturbative nature of the propagators in Euclidean spacetime but accessing the entire 
complex plane from the Euclidean momenta and looking at the pole structure and branch cuts is challenging. The attempts to
extend the lattice results to the complex plane have used Pad\'e approximants and suggested the presence of complex
conjugate poles for the gluon propagator~\cite{Falcao:2020vyr,Binosi:2019ecz,Boito:2022rad} but not for the ghost~\cite{Falcao:2020vyr,Binosi:2019ecz}
or the quark~\cite{Falcao:2022gxt} propagators. For the ghost propagator the Pad\'e analysis suggest the presence of
a pole at zero momentum, while for the quark propagator the same procedure identifies a pole at Minkowski momenta that
is correlated with the pion mass.

The use of the Dyson-Schwinger equations (DSE)~\cite{Roberts:1994dr,Bashir:2012fs} 
allows to access the propagators for complex momenta~\cite{Maris:1994ux,Maris:1995ns}.
Although DSEs are a powerful tool, any solution has to rely on a truncated set of equations
and, typically, model some of the Green functions~\cite{Krein:1990sf,Maris:1996zg,Bender:1994bv,Alkofer:2003jj,Windisch:2016iud,Hayashi:2020myk}.
This explains, in part, the different results that identify complex conjugate poles and/or simple poles for some of the propagators.
The combination of the DSE with spectral methods~\cite{Salam:1963sa,Salam:1964zk} offers another way to
explore the analytic structure of the propagators but it is best suited for QED rather than for QCD, due to the complexities associated
with the non-Abelian nature of QCD. However, despite its limitations to investigate non-Abelian gauge theories, these combined techniques allow also to discuss the solutions of the DSE in Minkowski spacetime \cite{Jia:2017niz,Solis:2019fzm,Duarte:2022yur,Mezrag:2020iuo}.

Recently, there has
been an effort to measure the quark and gluon spectral densities from the Euclidean data directly or to understand, from first principles,
properties of the associated propagators~\cite{Dudal:2013yva,Lowdon:2017uqe,Fischer:2017kbq,Lowdon:2017gpp,Kades:2019wtd,Dudal:2019gvn,Li:2019hyv,Horak:2021pfr,Horak:2021syv,Dudal:2021gif,Li:2021wol,Horak:2022aza}. It is well known that the spectral function
fully determines the analytic structure of the propagator in Minkowski space.

In the current work the structure of the quark propagator in momentum space is investigated taking into account
the non-perturbative constraints of the QCD dynamics for the quark spectral densities derived in~\cite{Lowdon:2017gpp}. Following the suggestion of this work we assume that the quark propagator has a simple pole, that our analysis suggests to be at timelike momenta, 
together with other structures. Such contributions are essential to describe the quark propagator in the infrared (IR) region and are expressed effectively
by higher order poles, which are not associated with asymptotic states. 
A simple pole has also been seen in ~\cite{Falcao:2022gxt}, that relies on the analysis with
Pad\'e approximants, and can be interpreted as giving supports to  
constituent quark models. The extra structures beyond the simple pole that, herein, are parametrized effectively as higher order 
poles,
are essential to the description of the quark
propagator
and should be responsible for the quarks
positivity violation, which has been used many times as an indication of quark confinement.

Our starting point, following \cite{Lowdon:2017gpp},  is that the spectral densities associated with the scalar and vector component
of the quark propagator, share a common simple pole but not its residuum. The lattice data favours a simple pole at timelike momenta, that can be associated with a typical constituent quark mass. In addition, the spectral density includes functionals, for example, derivatives of the Dirac-delta lead to higher order poles that do not correspond to asymptotic physical states. Furthermore, in our analysis the poles of the quark propagator can appear as zeros of the quark wave function. However,
not all zeros of the quark wave function translate into poles in the propagator.

Asymptotic scaling laws are derived from the effective parametrization of the quark propagator beyond the simple pole. These scaling laws are tested against lattice data for the Landau gauge quark propagator by using functional forms that they motivate.
It turns out that these functional forms are able to reproduce quite well the lattice data over a wide range of momenta and, in this sense,
they are validated by the lattice data.
Finally, the pion Bethe-Salpeter vertex and amplitude, in the chiral limit, are explored relying on the results derived herein. This allows 
to estimate the mass associated with the shared pole, estimate the pion Bethe-Salpeter amplitude from the lattice data combined with the
functional forms and estimate the pion decay constant from our analysis. Our analysis returns a pole mass for the shared pole 
that reproduces a typical  constituent quark mass of about 300 MeV. Again, this result can be viewed as given support to constituent quark models. 
Moreover, relying on the modelling of the lattice data discussed in this manuscript, the pion decay constant is computed and we find a value quite close to the experimental value that support the adopted description of the lattice data.

This work is organized as follows. In Sect. 2, we propose the analytical form of the quark propagator, separating the simple pole and the contributions from higher order poles, as well as the constraints on the ultraviolet behaviour derived from the proposed form. In Sect. 3, we suggest an analytical form for the quark mass and wave functions. In Sect. 4,  a suggested parametrization of the running quark mass and wave function based on the discussion of Sect. 3, is shown to describe well the Lattice results for the Landau Gauge. In Sect. 5, based on the axial-vector Ward identity in the chiral limit we parameterize the pion axial vertex and check the consistence of our parametrizations by estimating the position of the timelike pole, as suggested in Ref. \cite{Lowdon:2017gpp} corroborated by the evidences found in Ref.\cite{Falcao:2022gxt} from Padé analysis of the quark propagator.We also compute the pion decay constant. We close our work with a summary of our results in Sect. 6.

\section{On the quark propagator analytical structure \label{Sec:Theo}}

In momentum space the Euclidean quark propagator can be written as
\begin{equation}
    S(p) = - i \, \sigma_V(p^2) \, \psla{p} ~ + ~ \sigma_S(p^2) = Z(p^2) ~ \frac{ - i \, \psla{p} + M(p^2)}{p^2 + M^2(p^2)}\, , 
\end{equation}
where $\sigma_V(p^2)$ and $\sigma_S(p^2)$ are Lorentz scalar form factors, $Z(p^2)$ is the quark wave function and $M(p^2)$ the renormalization 
group invariant running quark mass. Assuming that the quark propagator has a spectral representation\footnote{
The quark spectral representation can also be viewed as an integral representation of the propagator.}, and following~\cite{Lowdon:2017gpp}
 that suggests that the form factors $\sigma_V(p^2)$ and $\sigma_S(p^2)$ share a common
simple pole mass term~\cite{Lowdon:2017gpp} it follows that one can write
\begin{eqnarray}
    \sigma_S(p^2) & = & \frac{Z_S}{p^2 + m^2} + \Sigma_S(p^2) \ , \label{Eq_SigmaS} \\
    \sigma_V(p^2) & = & \frac{Z_V}{p^2 + m^2} + \Sigma_V(p^2) \,,\label{Eq_SigmaV} 
\end{eqnarray}
where $m$ is the common pole mass, $Z_S$ and $Z_V$ are the residua of the scalar and vector components of the propagator at the pole mass, respectively, 
and $\Sigma_S$, $\Sigma_V$ summarize all the remaining structures of the propagator. fIt follows from the definitions that $Z_V$ is dimensionless, $Z_S$ has dimensions of $mass$, $\Sigma_S$ has dimensions of $1/mass$ and 
$\Sigma_V$ dimensions of $1/mass^{2}$. The decomposition in Eqs.~(\ref{Eq_SigmaS}) and (\ref{Eq_SigmaV}) constrain the support of the quark propagator onto the forward light-cone. 
A possible effective representation for the $\Sigma_{v,s}$ is an infinite series of higher-poles that also helps to understand its UV behaviour. It is well known that in the UV there are $\log$ corrections that we do not take into account. 
Note that given the precision of the lattice data and the range of momenta accessed by the simulations it would be impossible to identify the
$\log$ corrections.

The asymptotic expressions for the propagator are, up to log corrections,
\begin{equation}
    \sigma_V(p^2) \rightarrow \frac{1}{p^2} \qquad\mbox{ and }\qquad \sigma_S(p^2) \rightarrow \frac{1}{p^2} \ .
\end{equation}
Then, the allowed asymptotic behaviour for $\Sigma_{V,S}$ is
\begin{equation}
    \Sigma_V(p^2) \rightarrow \frac{1}{\left( p^2 \right)^{1 + \delta_V}} \quad\mbox{ and }\quad \Sigma_S(p^2) \rightarrow \frac{1}{\left( p^2 \right)^{1 + \delta_S}}
    \label{SigmaAsymptoticsOriginal}
\end{equation}
where the real positive numbers are such that $\delta_V \geq 0$ and $\delta_S \geq 0$.

In the following it will be assumed, as suggest by~\cite{Lowdon:2017gpp}, the pole propagator at $m^2$ is a simple pole and, therefore, 
that $\Sigma_S$ and $\Sigma_V$ are regular at the pole. The quark running mass and the quark wave function can be rebuilt 
from the functions $\sigma_S$ and $\sigma_V$ and are given by
\begin{eqnarray}
    M(p^2) & = & \frac{\sigma_S (p^2) }{\sigma_V(p^2)} = \frac{Z_S + \left( p^2 + m^2 \right) \Sigma_S(p^2)}{Z_V + \left( p^2 + m^2 \right) \Sigma_V(p^2)}\ , 
               \label{Eq:M1} \\
    Z(p^2) & = & \sigma_V(p^2) \, \left( p^2 + M^2(p^2) \right) \nonumber \\
    & = & \Bigg( Z_V + \left( p^2 + m^2 \right) \Sigma_V(p^2) \Bigg) \frac{ p^2   + M^2(p^2)}{p^2 + m^2} .
                \label{Eq:Z1}
\end{eqnarray}
As can be read from Eqs.~(\ref{Eq:M1}) and (\ref{Eq:Z1}) the information on the pole shared by the scalar and vector part of the propagator
does not appear in association with the running mass but with the quark wave function. Furthermore, any pole observed in the running mass
will also be seen in the quark wave function, unless 
$Z_V + \left( p^2 + m^2 \right) \Sigma_V(p^2)$ or $p^2+M^2(p^2)$ are zero at exactly the same momentum.

In the UV regime, due to asymptotic freedom, the running quark mass and the quark wave function are determined by the perturbative solution of QCD. 
For large momentum 
\begin{eqnarray}
      M(p^2) & = &  \frac{Z_S + p^2 \Sigma_S(p^2)}{Z_V + p^2 \Sigma_V(p^2)} \ , \\
      Z(p^2) & = & \Bigg( Z_V + p^2 \, \Sigma_V(p^2)\Bigg) ~~ \Bigg[ 1   + \frac{M^2(p^2)}{p^2}  \Bigg] \ , \label{Eq:Zasymptotic}
\end{eqnarray}
with the perturbative solution, in the Landau gauge, being
\begin{eqnarray}
      M(p^2) & = & \frac{\bar{m}}{\left( \log\frac{p^2}{\mu^2} \right)^{ \gamma_m} } \ ,\\
      Z(p^2) & = & \mbox{Constant} \, ,\label{Eq:ZAs}
\end{eqnarray}
where $\bar{m}$ is a mass scale, $\gamma_m > 0$ is the quark anomalous dimension and $\mu$ is another mass scale. Combining the asymptotic
expressions for the masses, it comes that
\begin{equation}
 \Sigma_S(p^2) = \Bigg( \frac{Z_V}{p^2}  +  \Sigma_V(p^2) \Bigg) \, \frac{\bar{m}}{\left( \log\frac{p^2}{\mu^2} \right)^{ \gamma_m} } - \frac{Z_S}{p^2}
\end{equation}
that combined with Eq.~(\ref{SigmaAsymptoticsOriginal}) implies that $\Sigma_S(p^2)$ vanishes in the UV limit and, therefore, $\delta_S > 0$. 
The role of the log correction is fundamental to ensure this result.

In order to proceed let us write 
\begin{eqnarray}
    \Sigma_S(p^2) & = &  \left( p^2 \right)^\beta \, \sum_{i=s}^{+ \infty} \frac{b_i}{\left( p^2 + \widetilde{m}_i^2 \right)^i}\, , 
    \label{Eq:SeriesS} \\
    \Sigma_V(p^2) & = &  \left( p^2 \right)^\alpha \, \sum_{i=s^\prime}^{+ \infty} \frac{a_i}{\left( p^2 + m_i^2 \right)^i}\, ,
    \label{Eq:SeriesV}
\end{eqnarray}
where $\alpha$ and $\beta$ are real numbers, $s$ and $s^\prime$ are integers, $a_i$ and $b_i$ are real numbers that determine the two functions and
$\widetilde{m}_i$ and $m_i$ are the mass scales that characterize $\Sigma_S$ and $\Sigma_V$, respectively.
We are, for the momenta, ignoring possible log corrections that would need to be regularized to have a smooth function for the full range of momenta.
The series have to be compatible with the asymptotic UV behaviour and, therefore,
\begin{equation}
    s - \beta -1 > 0 \qquad\mbox{ and }\qquad s^\prime - \alpha -1 \geq 0
    \label{Restricao}
\end{equation}
For the running mass this implies
\begin{equation}
     M(p^2)  =   \frac{Z_S + b_s \, \left( p^2 \right)^{\beta - s + 1} }{Z_V + a_{s^\prime} \left( p^2 \right)^{\alpha - s^\prime +1}} 
     = \mbox{ Constant} \ .
\end{equation}
This equation has two types of solutions. One class of solutions has
\begin{equation}
     \beta = s - 1 \qquad\mbox{ and }\qquad \alpha = s^\prime -1 \ ,
\end{equation}
and $\beta$ and $\alpha$ are integer numbers, while the second class of solutions has
\begin{equation}
     \beta - s = \alpha - s^\prime \, ,
\end{equation}
with $b_s$ being related to $a_{s^\prime}$ and the constant itself.
The first class of solution does not comply with the constrain~(\ref{Restricao}) and, therefore, it should be disregarded.
For the second class of solutions the constrains in Eq.~(\ref{Restricao}) become
$s^\prime - \alpha -1 > 0$. The series can now be written as
\begin{eqnarray}
    \Sigma_S(p^2) & = &  \left( p^2 \right)^{\alpha + s - s^\prime} \, \sum_{i=0}^{+ \infty} \frac{b_i}{\left( p^2 + \widetilde{m}_i^2 \right)^{i+s}} \, ,
    \label{SigmaS}\\
    \Sigma_V(p^2) & = &  \left( p^2 \right)^\alpha \, \sum_{i=0}^{+ \infty} \frac{a_i}{\left( p^2 + m_i^2 \right)^{i+s^\prime}}\, .
    \label{SigmaV}
\end{eqnarray}
If these functions are regular for all momenta including the zero momentum, then $\alpha \geq 0$,
$\alpha + s - s^\prime \geq 0$ and $s^\prime > \alpha +1$. 
The lowest value for the exponent of $\Sigma_V$ is $\alpha = 0$ that implies
$s \geq s^\prime > 1$ and, in this case, the two series can start with a double pole or a higher order pole. 
Other solutions with higher order poles are also possible, with higher values for $\alpha$ requiring larger $s^\prime$ and $s$.

At zero momentum, the running mass and the wave function read
\begin{eqnarray}
    M(0) & = & \frac{ Z_S + m^2 \, \Sigma_S(0)}{Z_V + m^2 \, \Sigma_V(0)} \, ,\\
    Z(0) & = & \frac{M^2(0)}{m^2} \, \Big( Z_V + m^2 \, \Sigma_V(0) \Big) \, ,
\end{eqnarray}
and, therefore, $Z(0)$, the pole mass and the IR mass scale, i.e. the zero momentum running mass,
are not independent numbers. By plugging in typical numbers from lattice simulations results that give $Z(0) \sim 0.8$, $M(0) \sim 0.35\,$GeV ~\cite{Oliveira:2018lln} and $m \sim 0.3\,$GeV~\cite{Falcao:2022gxt} it
is possible to estimate the ratio of the term in parenthesis
\begin{equation}
    Z_V + m^2 \, \Sigma_V(0) \sim 0.2 \ .
    \label{EqEstimativa}
\end{equation}
If $\alpha > 0$ and $\alpha + s - s^\prime > 0$, then Eqs.~\eqref{SigmaS} and \eqref{SigmaV} give $\Sigma_S(0) = \Sigma_V(0) = 0$ and the expressions for
$M(0)$ and $Z(0)$ become simpler
\begin{eqnarray}
    M(0)  =  \frac{ Z_S }{Z_V } \qquad\mbox{ and }\qquad 
    Z(0)  =  \frac{Z^2_S}{Z_V \, m^2} \ .
\end{eqnarray}

\section{The pole structure}

For convenience, let us redefine $\Sigma_S = Z_S \, \Sigma^\prime_S $ and $\Sigma_V = Z_V \, \Sigma^\prime_V$ such that: 
\begin{eqnarray}
    M(p^2) & = & \frac{Z_S}{Z_V} \,  \frac{1 + \left( p^2 + m^2 \right) \Sigma_S(p^2)}{1 + \left( p^2 + m^2 \right) \Sigma_V(p^2)}\ , 
               \label{Eq:M2} \\
    Z(p^2) & = & Z_V \, \Bigg( 1 + \left( p^2 + m^2 \right) \Sigma_V(p^2)\Bigg) \, \times
    \nonumber \\
    &  & \qquad \times \, \frac{ p^2   + M^2(p^2) }{p^2 + m^2} \, ,
                \label{Eq:Z2}
\end{eqnarray}
where we have omitted the prime in the above expressions to simplify the notation. In the series (\ref{SigmaS}) and (\ref{SigmaV})
this procedure is equivalent to
replace $b_i \rightarrow Z_S \, b_i$ and $a_i \rightarrow Z_V \, a_i$. The rescaling of the $\Sigma_{S,V}$ by $Z_{S,V}$,
respectively, changes the units of the original quantities and the $\Sigma_{S,V}$ have now dimensions of $1/mass^2$, as can be read
from Eq.~(\ref{Eq:Z2}). Recall that $Z_S$ has dimensions of $mass$, while $Z_V$ is dimensionless.

The quark wave function has a pole at the timelike momenta $p^2 = - m^2$. However, this pole does not show up in $Z(p^2)$ if
\begin{equation}
     m^2 = M^2(-m^2) = \frac{Z^2_S}{Z^2_V}\, , \label{GapEq}
\end{equation}
and, therefore, it appears as pole of the quark propagator itself. Note also that, in this case,
the quark wave function is
\begin{equation}
        Z(p^2=-m^2)  =  Z_V \, .
\end{equation}
Moreover, if (\ref{GapEq}) has more than one solution for $p^2$, then the quark wave function should have a zero for each of the
$p^2$ that solve this equation as there are no terms in the denominator that that can compensate the zero. The other possibility 
to compensate the would be simple pole in the propagator is to have a singular behavior, near the pole, of $\Sigma_V$.
If $Z(p^2)$ has no zeros, then Eq.~(\ref{GapEq}) should have a unique solution and the quark propagator has a unique simple 
pole at the timelike momenta $p^2 = - m^2$. Surprisingly, this was exactly what was found in the analysis of the analytic structure of the Landau gauge
lattice quark propagator data using Pad\'e approximants in~\cite{Falcao:2022gxt}.

\section{The Lattice data}

Let us investigate if the results summarized in Eqs. (\ref{Eq:M2}) and (\ref{Eq:Z2}) are compatible with the outcome of the
lattice simulations for the quark propagator in the Landau gauge. For the lattice data we will consider
that reported in~\cite{Oliveira:2018lln} and, of the various simulations, only that
performed with a $\beta = 5.29$ that use the $64^4$ lattice will be analysed. The measured pion mass in this simulation is 290 MeV. 
The rationale for choosing this lattice data is to consider the simulation with a larger number of points in the infrared region
and a good statistics\footnote{The work \cite{Oliveira:2018lln} computes the quark propagator for various ensembles. Overall, the propagators are qualitatively similar, with small differences at low momenta that should not impact our study.}. The available lattice results are for the running quark mass $M(p^2)$ and for 
the quark wave function $Z(p^2)$.  The direct use of Eqs.~(\ref{Eq:M2}) and (\ref{Eq:Z2}) with the simplest possible
expressions for $\Sigma_{V,S}$ is a bit cumbersome, as the corresponding fits are hard to perform. So instead, we choose
to fit directly $M(p^2)$ and $Z(p^2)$ assuming functional forms motivated by the results derived at the end of Sec.~\ref{Sec:Theo} that tell that
the series for $\Sigma_{S,V}$ start with a $\sim 1/(p^2 + m^2)^2$ type term.

\begin{figure}
    \centering
    \includegraphics[width=1.1\linewidth]{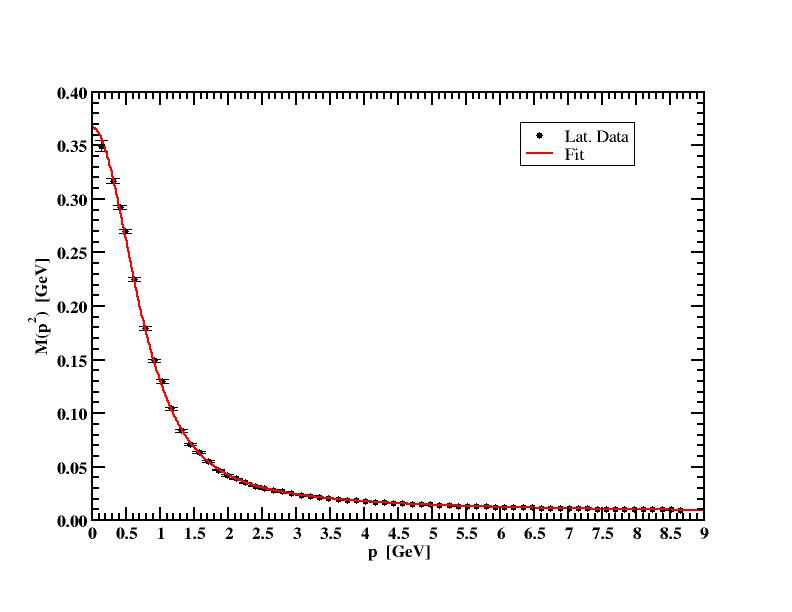} \\
    \includegraphics[width=1.1\linewidth]{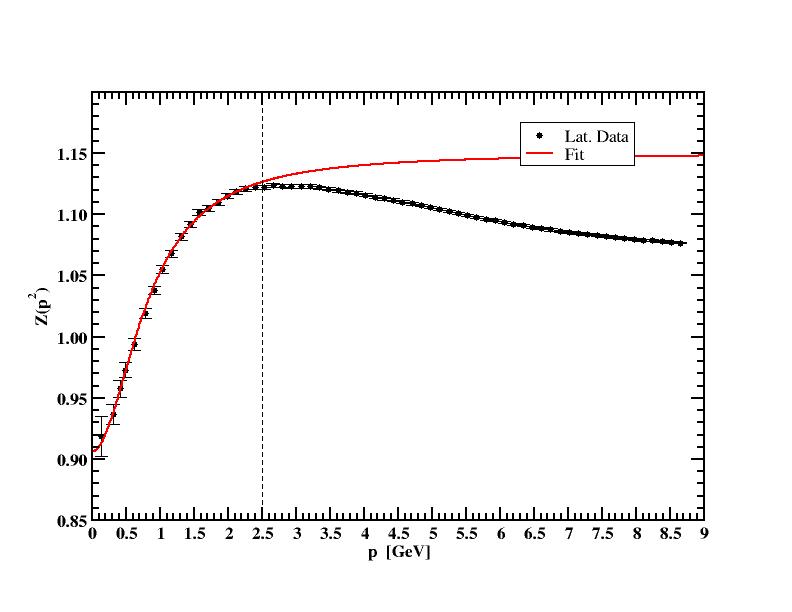} \\
    \caption{Running quark mass (top) and quark wave function (bottom) from the lattice simulations and the fitted functions.
            Set main text for details. In the fit to the quark wave function only the lattice data up to $p = 2.5$ GeV (dashed
            black line) was considered.}
    \label{fig:running-mass}
\end{figure}

In order to find suitable functional forms compatible with the lattice data, let us assume that $\Sigma_V \sim 0$. Then, it follows
from Eq.~(\ref{Eq:M2}) that the running mass reads
\begin{eqnarray}
 M(p^2) & \approx &  \frac{Z_S}{Z_V} \Bigg( 1 + (p^2 + m^2) \Sigma_S(p^2) \Bigg) \nonumber \\
        & \approx &  \frac{Z_S}{Z_V} \left( 1 + \frac{b_1}{p^2 + m^2_1} + \frac{b_2}{\left( p^2 + m^2_2 \right)^2} + \cdots \right) \, .
\end{eqnarray}
Furthermore, inserting this expansion in Eq.~(\ref{Eq:Z2}) one arrives at the same type of expansion for the quark wave function.
Therefore, for the running quark mass it will be assumed that
\begin{equation}
    M(p^2) = m_0 + \frac{a_1}{p^2 +  b_1} + \frac{a_2}{(p^2 + b_2)^2} \ .
    \label{Eq:Fitmass}
\end{equation}
For $a_2 = 0$ this expression reproduces the suggestion of~\cite{Cornwall:1981zr} for the running gluon mass
and often used to parameterise the lattice data or the solutions of the Dyson-Schwinger equations, see e.g.~\cite{Aguilar:2019uob,Oliveira:2018ukh,Castro:2023bij,Moita:2022lfu} and references therein. 
The parameter $m_0$ can be interpreted as the current quark mass that should be recovered in the UV regime. Indeed, as can be seen below,
$m_0 \sim 7 $ MeV, in good agreement with this view.
The fit with $a_2 = 0$ results in a curve that is not
that far away from the lattice data, but whose $\chi^2/d.o.f. \sim 16$. On the other hand, using all the three terms results
in the fit seen in Fig.~\ref{Eq:Fitmass} that has a $\chi^2/d.o.f. = 0.77$. In this case, the fitted parameters are
$m_0 = 0.007003(42)\,$GeV, $a_1 = 0.2403(40)$\,GeV$^3$, $b_1 = 10.04(41)$\,GeV$^2$, \\ $a_2 = 0.506(11)$\,GeV$^5$ and $b_2 = 1.227 (14)$\,GeV$^2$. Notice that the fits do not account for correlations.

For the quark wave function it was assumed that
\begin{equation}
    Z(p^2) = z_0 + \frac{z_1}{p^2 + m^2_1}\, ,
    \label{Eq:FitZ}
\end{equation}
and the lattice data was considered up to $p = 2.5$\,GeV. Above this value of the momentum the lattice $Z(p^2)$ decreases as $p$
increases, see bottom plot in Fig.~\ref{fig:running-mass}, and this is considered to be a finite lattice spacing effect - see the discussions in \cite{Oliveira:2018lln}. The fit to the lattice data results in a $\chi^2/d.o.f. = 0.49$ and
$z_0 = 1.1500(15)$, $z_1 = - 0.1639(69)$\,GeV$^2$ and $m^2_1 = 0.672(35)$\,GeV$^2$. 

We recall the reader that we are using an effective parameterization of the self energies. More, we remind that the lattice data has
a finite precision and covers a limited range of momenta. Hopefully, the leading term in the expansion will be well determined. Anyway, the parameters found express the UV and IR regions. In that respect, the running quark mass has a simple pole, with a mass scale of $\sim \sqrt{b_1} \sim 3\,$GeV that dominates the UV region, which is expected to be much larger than $\Lambda_{QCD}$ (one order of magnitude) and consistently contribute with less than 10\% to the infrared mass. The double pole in the running mass is associated with double poles in the quark propagator of non-perturbative origin, and it should carry a typical non-perturbative or IR  mass scale of $\mathcal{O}(\Lambda_{QCD})$, which in our case $\sqrt{b_2} \sim 1$ GeV. The quark renormalization has a simple pole $m_1 \sim 1$ \,GeV representing also the non-perturbative QCD physics. Note that a similar reasoning for the gluon propagator, where the spectral density includes derivatives of delta functions that translates into higher order poles, the mass scale associated to the double pole term was found to be also of $\sim 1 $GeV, a typical non perturbative scale \cite{Li:2019hyv}.
The important message in our  work is that the functional forms suggested by the theoretical analysis of Ref.~\cite{Lowdon:2017gpp} describes extremely well the chosen set of lattice QCD results.

The conclusion of this section being that the results of Sec. \ref{Sec:Theo} do not only offer a 
picture for the quark propagator, they are also compatible with the lattice results for the Landau gauge quark propagator. In addition, we should emphasize to the reader the importance of the lattice results in the IR region to fit the double pole in the mass function as well as the simple pole in the quark wave function renormalization.

\section{Phenomenological implications: the pion Bethe-Salpeter amplitude}

\begin{figure}
    \centering
    \includegraphics[width=1.1\linewidth]{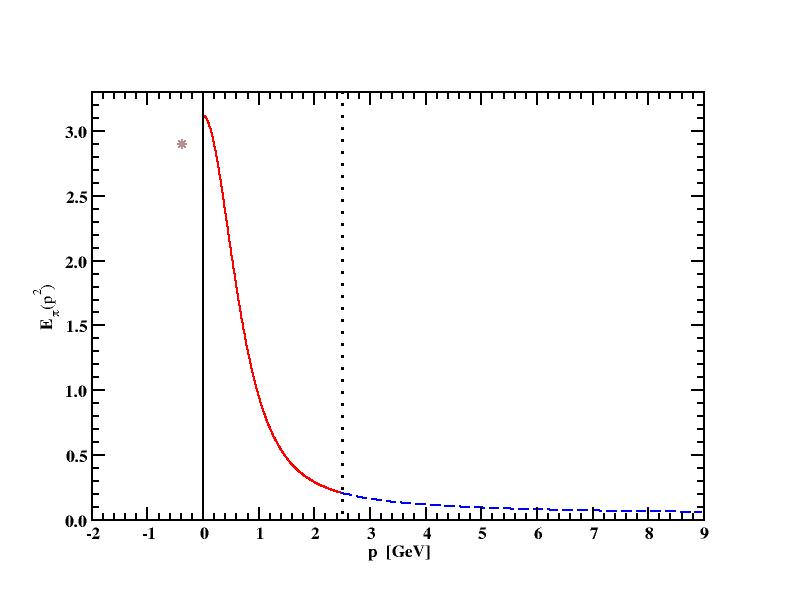}
    \caption{Pion pseudoscalar vertex in the chiral limit, $E_{\pi} (p^2)$, for Euclidean momentum derived from Eq.~\eqref{Eq:EMZ} based on our fits of $M(p^2)$ and $Z(p^2)$ taking into account the experimental value of $f_\pi=130\,$MeV~\cite{Workman:2022ynf}.}
    \label{fig:Epi_P}
\end{figure}

We can apply the analysis presented in this work to compute pion properties. The pseudoscalar Bethe-Salpeter (BS) vertex, $\Gamma_\pi(p,P)$, can be written as (see e.g.~\cite{QCD50Y})
\begin{eqnarray}
\label{vertex}
  && \Gamma_\pi (p,P) = \gamma_5 [i \,  E_\pi (p,P)+ \psla{P}  \, F_\pi (p,P) \nonumber
\\&& \qquad + ~  p^\mu P_\mu\,  \psla{p} \, G_\pi (p,P) + \sigma_{\mu\nu} p^\mu P^\nu H_\pi (p,P)]\, ,
\end{eqnarray}
where $p$ is the relative momentum of the constituents, $P$ is the total bound state momentum and 
$E_\pi(p,P)$, $ F_\pi (p;P) $, $ G_\pi (p;P) $ and $H_\pi (p,P)$ are four scalar amplitudes. 
In the chiral limit, we consider that the contribution to the BS vertex comes from $E_\pi(p,0) = E_\pi(p)$, which is given by
\cite{Maris:1997hd} (see also~~\cite{QCD50Y})
\begin{equation}
    f_\pi \, E_{\pi} (p) = M(p^2)/Z(p^2) \, .
    \label{Eq:EMZ}
\end{equation}

The function $E_{\pi} (p)$ can be estimated from the lattice data for the Landau gauge quark propagator. Using the fits discussed in the previous section,
it turns out that $E_{\pi} (p)$ is as given in Fig.~\ref{fig:Epi_P}, where the solid red line is the results from the fits up to $p = 2.5$\,GeV and
the dashed blue line its extension up to $p = 9$\,GeV. The dotted vertical line signals the $p = 2.5$\,GeV and refers to the maximum momentum used in
the fit to the quark wave function. 

From the fits to the lattice data quark propagator one estimates $E_\pi(0) = 3.12$. Then, combining Eq.~(\ref{Eq:Z2}) with (\ref{Eq:EMZ})
it follows that
\begin{equation}
    f_\pi \, E_{\pi} (0) = \frac{m^2}{Z_V \Big( 1 + m^2 \, \Sigma_V(0) \Big) \, M(0)} \, ,
    \label{Eq:EMZ2}
\end{equation}
that allows to estimate the pole mass $m$. Using $f_\pi = 130$\,MeV, $M(0) = 350$\,MeV it comes
out $m  \sim 377$\,MeV for $Z_V \, ( 1 + m^2 \, \Sigma_V(0) ) \sim 1$, that is  within the typical value 
associated with the constituent quark mass and is a value close to $M(0)$. 
Once having an estimation of the pole mass, after going back to Eq.~(\ref{Eq:EMZ}) it is possible to estimate $E_\pi$ 
at the pole mass that is $E_\pi( \sqrt{- m^2}) = 2.9/Z_V$. For a $Z_V = 1$ this point is represented in Fig.~\ref{fig:Epi_P} by the brown star at negative
$p$. 

In addition, we can write the Bethe-Salpeter amplitude (BSA), in the chiral limit, as
\begin{equation}
    \Psi_{\pi}(p, P) = S\bigg(p+\frac P2 \bigg) \, i \, \gamma^5 \, E_\pi(p)\, S\bigg(p-\frac P2\bigg)\, ,
    \label{BSA}
\end{equation}
where $S$ is the quark propagator and the momenta are the Minkowski ones.
In terms of the BSA, the pion decay constant can be written as follows \cite{dePaula:2020qna}
\begin{equation}
i~p^\mu f_{\pi} = N_{c} \int \frac{d^4 k}{ (2\pi)^4}~\mbox{Tr} [\, \gamma^\mu \, \gamma^5 \, \Psi_\pi(k ,p) ] \, .
\label{fpi0}
\end{equation}
Note that with this definition $f_\pi = 130/\sqrt{2} = 92$ MeV.
Using the fit for the running quark mass, Eq.~(\ref{Eq:Fitmass}), and for the quark wave function, 
Eq.~(\ref{Eq:FitZ}), one can calculate $f_{\pi}$. For this end, we have to combine Eqs.~(\ref{Eq:EMZ}) using the relation 
$E_\pi(p)=M(p^2)/(f_\pi Z(p^2))$, Eq. (\ref{BSA}) and leaving in Eq.~(\ref{fpi0}) $f_\pi$ as the unknown. For $N_c = 3$, the result is $f_{\pi} = 95$ MeV.

We clarify that our fitting of the running quark mass and $Z(p^2)$  do not use the pion decay constant. So, including the other terms in the pion-quark-antiquark vertex will not affect our numerical analysis. However, we emphasize that in our estimation of $f_\pi$ we used only the $\gamma_5$ pion vertex, which of course is not all the components as the reader can notice, but it gives a fair reproduction of pion observables as, for example, the parton distribution function as shown with the algebraic model given by Ref. \cite{Chang:2013pq}. For this reason, we have adopted the single $\gamma_5$ vertex.

\section{Summary and Conclusions}

In the current work the momentum space structure of the quark propagator is studied taking into account the QCD non-perturbative constraints for the
spectral densities derived in Ref.~\cite{Lowdon:2017gpp}, that claims that the scalar and vector components of the quark propagator 
share a common timelike simple pole but not its residuum. For the remaining part of the propagator but the simple pole, we assumed that it can
be effectively described via higher order poles, similarly to the procedure used to describe the gluon propagator in~\cite{Lowdon:2018mbn}. 
Indeed, for the gluon propagator in the Landau gauge it was found that the lattice results in the infrared region are well 
fitted by a simple and a double pole contribution~\cite{Lowdon:2017gpp}. 
However, its spectral density may have higher order functionals, like derivatives of the Dirac-delta function, that can be understood as a 
manifestation of color confinement~\cite{Lowdon:2018mbn}. 
 
The above assumptions allowed us to establish a connection between the poles of the quark propagator and the zeros of the quark wave function $Z(p^2)$.
The analytic structure that was built suggests that not all the zeros of $Z(p^2)$ are necessarily associated with poles in the propagator. 
On the top of the general analytic form, asymptotic scaling laws for the representation of the quark propagator were derived, after removing the simple pole contribution shared by the scalar and vector parts. 

The general analytical form for the quark propagator containing a simple timelike pole, shared by the scalar and vector parts, added by
higher order poles suggests simple parametrizations for the running quark mass and the quark wave function. For the mass function it 
contains a single and a double pole. The simple pole was suggested long ago in Ref.~\cite{Cornwall:1981zr} for the gluon running mass and 
for the running quark quark mass to build the pion Bethe-Salpeter amplitude in~\cite{Mello:2017mor,Moita:2022lfu,Castro:2023bij}. 
On the other hand, for the quark wave function in the IR region, at least up to momenta $\sim 2.5$ GeV, it can be described by
a constant plus a simple pole term.
The confrontation of these functional forms, for the running mass and the quark wave function, with the lattice data for the full QCD quark propagator 
are in good agreement. Motivated by that an interesting point  to be studied in the future is the search for a kernel of the Schwinger-Dyson equation compatible with the integral representation of the quark propagator on the forward light-cone, with higher order poles.

Finally, by exploring the link between the quark self-energy and the pion vertex function in the chiral limit, 
as provided by the axial-vector Ward-Takahashi identity (see e.g.~\cite{QCD50Y}) we estimated the position of the shared simple timelike pole at $m\sim 377\,$MeV, to be compared with 300 MeV obtained from a Pad\'e analysis of the quark propagator in Landau gauge \cite{Falcao:2022gxt},  that is of the order of a typical constituent quark mass found in the literature. A cross-check of the proposed 
parametrization for the quark propagator and the suggested pion Bethe-Salpeter amplitude, built from the quark self-energies and the axial vertex, 
is the computation of the pion decay constant, which was found to be consistent with the experimental data. 
The results found here motivate us to, in the future, explore the implications of the suggested parametrizations to compute the Minkowski space pion observables,  
  that are also of current experimental interest as, for example, in the exploration of the origin of 
the visible mass in the future Electron-Ion-Collider~\cite{Aguilar:2019teb,AbdulKhalek:2021gbh,Accardi:2023chb}.

\section*{Acknowledgments}

This work was partly supported by the FCT  Funda\c{c}\~ao para a Ci\^encia e a Tecnologia, I.P., under Projects Nos. 
UIDB/04564/2020 (\url{https://doi.org/10.54499/UIDB/04564/2020}), UIDP/04564/2020 (\url{https://doi.org/10.54499/UIDP/04564/2020}). O.O. also acknowledges financial support from grant 2022/05328-3, from S\~ao Paulo Research Foundation (FAPESP). T. F. thanks
the financial support from the Brazilian Institutions:
National Council for Scientific and Technological Development CNPq (Grant No. 308486/2015-3), Improvement of Higher Education Personnel CAPES (Finance Code 001) and FAPESP (Grants 
No. 2017/05660-0 and 2019/07767-1). W. d. P. acknowledges the partial support of CNPQ under Grant No. 313030/2021-9 and the partial support of CAPES under Grant No. 88881.309870/2018-01. This work is a part of the
project Instituto Nacional de  Ci\^{e}ncia e Tecnologia - F\'{\i}sica
Nuclear e Aplica\c{c}\~{o}es  Proc. No. 464898/2014-5.

\end{document}